\documentclass[letterpaper]{emulateapj}
\usepackage{graphicx}
\usepackage{amsmath,cases}
\usepackage{color}
\newcommand{\msun}{{M}_{\odot}}
\newcommand{\rsun}{{R}_{\odot}}

\shorttitle{GRB rate at very high redshift}
\shortauthors{KINUGAWA, HARIKANE \& ASANO.}

\begin{document}

	
	\title{Long gamma-ray burst rate at very high redshift}
	

	\author{Tomoya Kinugawa\altaffilmark{1}, Yuichi Harikane\altaffilmark{2}, and Katsuaki Asano\altaffilmark{2}}
	%
	%

\altaffiltext{1}{Department of Astronomy, Faculty of Science, The University of Tokyo, 7-3-1, Hongo, Bunkyo-ku, Tokyo, 113-8654 Japan}

\altaffiltext{2}{Institute for Cosmic Ray Research, The University of Tokyo, 5-1-5 Kashiwa-no-ha, Kashiwa City, Chiba, 277-8582, Japan}

\begin{abstract}
	Future missions for long gammma-ray burst (GRB) observations at high redshift  such as HiZ-GUNDAM and THESEUS will provide clue to the star formation history in our universe.
	In this paper focusing on high redshift ($z>8$) GRBs, we calculate the detection rate of long GRBs by future observations, considering both Population (Pop) I\&II stars and Pop III stars as GRB progenitors.
	For the Pop I\&II star formation rate (SFR), we adopt an up-to-date model of high-redshift SFR based on the halo mass function and  dark matter accretion rate obtained from cosmological simulations.	
	We show that the Pop I\&II GRB rate steeply decreases with redshift.
	This would rather enable us to detect the different type of GRBs, Pop III GRBs, at very high redshift.
	If 10\% or more Pop III stars die as an ultra-long GRB, the future missions would detect such GRBs in one year in spite of their low fluence.
	More luminous GRBs are expected from massive compact Pop III stars produced via the binary merger. In our conventional case, the detection rate of such luminous GRBs is $3-20~{\rm yr^{-1}}~(z>8)$.
	Those future observations contribute to revealing of the Pop III star formation history.
\end{abstract}
\section{introduction}
{The gravitational collapse of massive stars is considered as the trigger of the long gamma-ray burst (GRB), which is confirmed by observing long GRBs associated with hypernovae \citep{Galama1998,Hjorth2003,Melandri2014}.
However, the long GRB rate is not simply proportional to the star formation rate (SFR) \cite[e.g.][]{WP10,Lien2014,Lien2015}.
Long GRBs, which are the most luminous astrophysical phenomenon, are a powerful tool to reveal the high redshift star formation especially at $z\gtrsim10$.}
The long GRB at the highest redshift ever observed is GRB 090429B with a photometric redshift
$z\sim9.4$ \citep{Cucchiara2011}.
There are future plans of long GRB observations at high redshift  such as HiZ-GUNDAM \citep{Yoshida2016} and THESEUS \citep{YUAN10,Amati2018,Rossi2018}.
Those observational missions can probe the high redshift universe.
We have to identify what type of stars become long GRB progenitors to calculate the long GRB rate at high redshift.
At present, the prediction of the GRB rate at $z \gtrsim 10$ is difficult because of the lack of the observational knowledge of GRBs and the SFR at $z\gtrsim10$.

As the long GRB progenitor at high redshift, many authors have considered not only Population I and II (Pop I\&II) stars, but also Population III (Pop III) stars \cite[e.g.][]{L2002,Bromm2006,Belczynski2007,Campisi2011,de Souza2011,Toma2011,Toma2016,Ghirlanda2015,Burlon2016}.
Pop III stars are first stars formed from the primordial gas with no metal.
 Pop III stars are more massive stars than  Pop I\&II stars due to a lack of effective coolant such as  metal and dusts \citep{Omukai2005,Dayel2018}. 
 It has been suggested that two different Pop III star formation modes.
 The first generation of Pop III stars (Pop III.1) are formed from the primordial gas unaffected by the previous star formation, where the main coolant is H$_2$ molecule \citep{Tegmark1997,Abel2002,Bromm2002,Yoshida2006}.
 Recent studies suggest that the radiation feedback from the massive protostar leads to the typical mass of Pop III.1 of $\sim 40 ~\msun$ \citep{Hosokawa2011}. 
 The second generation Pop III stars (Pop III.2) are formed from the no metal gas ionized by the radiation from the previous star formation \citep{Johnson2006,McKee2008}.
 In such ionized gases, the hydrogen dueteride (HD) cooling is more effective than H$_2$ cooling, so that the typical mass ($\sim 20~\msun$) is slightly less than the mass of Pop III.1 \citep{Hosokawa2012}.
 
 Some simulations show that disk fragmentations frequently occur, which implies the existence of binary Pop III stars \citep{Saigo2004,Machida2008,Stacy2013,Susa2014}.
 Kinugawa et al. (2014) predicted detection of gravitational waves (GWs) from binary black hole (BH) mergers originated from Pop III stars. The first GW detection with LIGO, GW150914 \citep{Abbot2016}, was a 30+30 $\msun$ binary black hole merger, which supports the BH binary formation from Pop III stars. 
 
 The Pop III star formation rate has been studied with semi-analytical method or numerical simulations \cite[e.g.][]{de Souza2011, Jhonson2013}.
 The Thomson scattering optical depth for cosmic microwave background photons measured with Planck \citep{Planck2014,Planck2016a,Planck2016b} is lower than the previous values measured with WMAP \citep{Dunkley2009}. This provides tight constraints on the star formation history at high redshift \citep{Visbal2015,Hartwig2016,Inayoshi2016}.

{Since the mass distribution of Pop III stars is biased to heavier range than those for Pop I\&II stars, we can expect that Pop III stars tend to launch long GRBs easily.
\cite{de Souza2011} calculated the Pop III SFR using a semi-analytical method and estimated the Pop III GRB rate.
\cite{Yoon2012} calculated the rotating Pop III stellar evolution as a GRB progenitor model and showed that rotating massive Pop III stars experience the chemically homogeneous evolution and can launch GRB jets at the final stage of their evolution.
On the other hand, \cite{Nakauchi2012} showed that Pop III blue supergiant stars, which hold a massive hydrogen envelope, may give rise to a GRB with a duration of $\sim10^5$ s in the observer frame because of the long mass accretion phase. The peak luminosity of the Pop III ultra long GRBs was estimated as $\sim 5\times10^{50} ~\rm erg~ s^{-1}$.
For binary Pop III stars, \cite{Belczynski2007} discussed the tidal spin-up and the envelope ejection by the binary interaction and calculated the number of GRB progenitors based on a criterion from the angular momentum and amount of envelope.
}

In this paper focusing on high redshift ($>8$) GRBs, we calculate the detection rate of long GRBs by future observations, considering both Pop I\&II stars and Pop III stars as GRB progenitors.
The future detections of GRBs at very high redshift ($z \gtrsim 10$) will unveil the star formation history in the very early era.
For Pop I\&II stars, we adopt an up-to-date model of high-redshift SFR proposed by Harikane et al. (2018)  with the halo mass function and  dark matter accretion rate obtained from cosmological simulations in Ishiyama et al. (2015).

On the other hand, in the case of Pop III stars, we consider both ultra-long GRBs from massive stars with heavy envelope and classical GRBs from massive compact stars, which experienced the binary merger. 

The long GRB rate from Pop I\&II stars at high redshift is discussed in \S \ref{GRBfromPopIand II}.
In \S \ref{GRBfromPopIII}, we consider the long GRB rate from Pop III stars considering two cases: the ultra-long GRB and classical GRB progenitors. Our results are summarized in \S 4.

\section{GRB from Pop I\&II stars}\label{GRBfromPopIand II}
 
\subsection{Pop I\&II star formation rate}

\begin{figure}[!ht]
	\begin{center}
		\includegraphics[width=1.0\hsize]{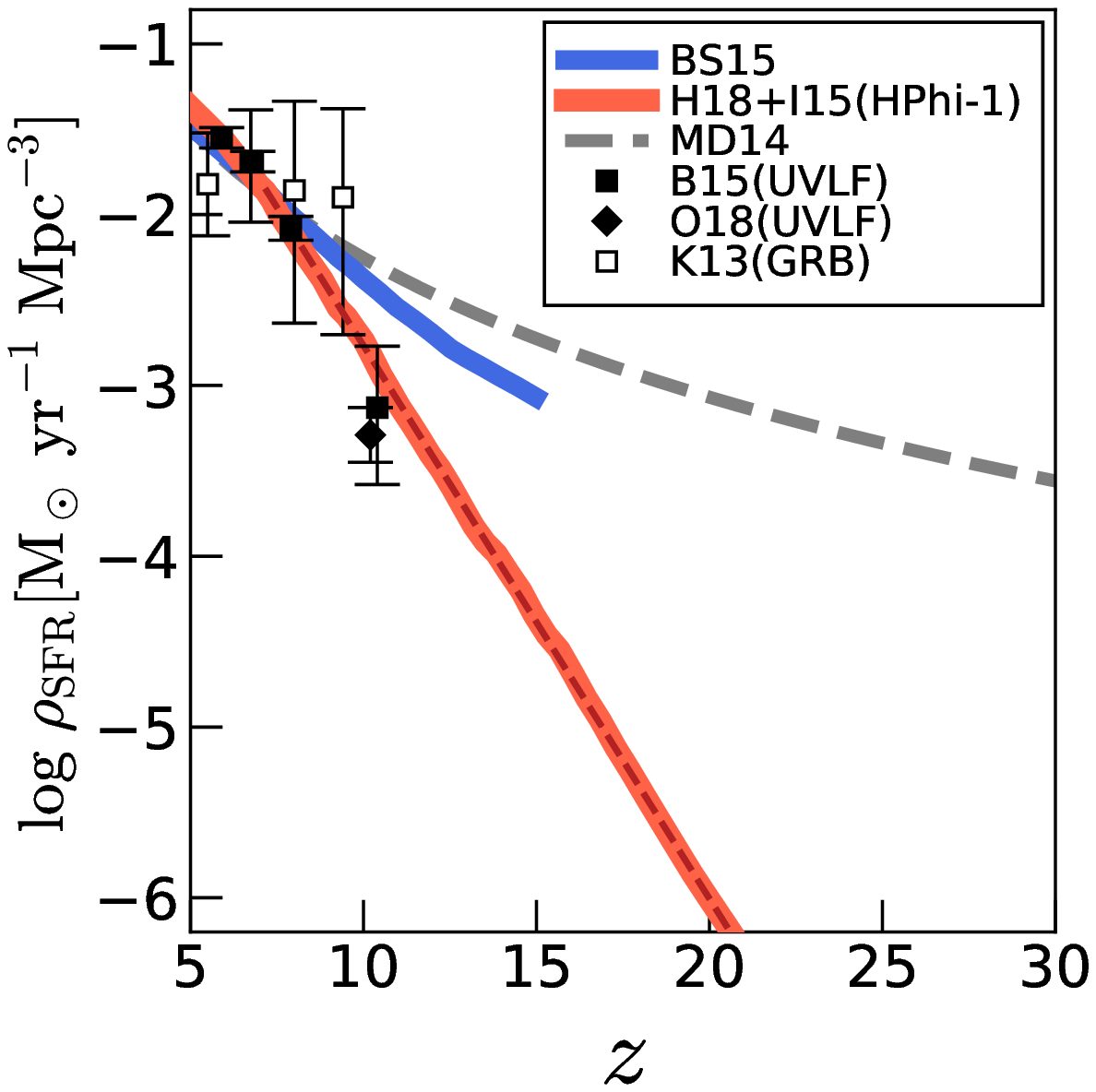}
	\end{center}
	\caption{Cosmic SFRD.
		The blue curve is the model calculations by \citet{2015ApJ...799...32B}.
		The red curve shows the cosmic SFRDs calculated based on the model in \citet{2018PASJ...70S..11H} with the N-body simulations in \citet{2015PASJ...67...61I} (the HPhi-1 model), and the red dashed line is the fitting function of Equation (\ref{eq_SFRDH18}).
		The dashed gray curve is the fitting function from \citet{2014ARA&A..52..415M} and its extrapolation.
		The black squares and diamond denote the observational results in \citet{2015ApJ...803...34B} and \citet{2018ApJ...855..105O}, respectively, based on the UV luminosity functions (UVLFs).
		The open squares show the SFRDs based on the long GRB observations in \citet{2013arXiv1305.1630K} normalized by \citet{2013ApJ...770...57B}.
		\label{fig_SFRD}}
\end{figure}

In order to calculate the long GRB rate from Pop I\&II stars at high redshift, first we estimate the cosmic star formation rate densities (SFRDs) of Pop I \& II stars.
Since the cosmic SFRDs at $z>10$ is poorly constrained from observations, we consider two models in \citet{2015ApJ...799...32B} and \citet{2018PASJ...70S..11H} for SFRDs.
Below we shortly review their calculations.
See  \citet{2015ApJ...799...32B} and \citet{2018PASJ...70S..11H} for more details.

In both the models, the SFRD $\rho_{\rm SFR}$ can be calculated based on the following equation:
\begin{eqnarray}
\rho_\mathrm{SFR}&=&\int dM_\mathrm{h}\frac{dn}{dM_\mathrm{h}}SFR\\
&=&\int dM_\mathrm{h}\frac{dn}{dM_\mathrm{h}}\dot{M_\mathrm{h}}\frac{SFR}{\dot{M_\mathrm{h}}},
\end{eqnarray}
where $M_\mathrm{h}$, $\frac{dn}{dM_\mathrm{h}}$, and $\dot{M_\mathrm{h}}$ are dark matter halo mass, halo mass function, and dark matter accretion rate, respectively.
In \citet{2015ApJ...799...32B}, they obtain halo mass function and the dark matter accretion rate at $z=5-15$ from the {\it Bolshoi} N-body simulation \citep{2011ApJ...740..102K}.
The Bolshoi simulation is calculated in the redshift range of $z=0-80$ in a 250 $h^{-1}$ Mpc with the mass resolution of $1.9\times10^{8}\ M_\odot$.
The SFR per halo $SFR$ in \citet{2015ApJ...799...32B} is expressed as
\begin{equation}
SFR(t)=\frac{dM_*}{dt}=\frac{dM_*}{dM_\mathrm{h}}\dot{M_\mathrm{h}}=\frac{\alpha M_*}{M_\mathrm{h}}\dot{M_\mathrm{h}},
\end{equation}
where $M_*$ is the stellar mass, and $\alpha$ is the ratio of the specific star formation rate (sSFR) and the halo specific mass accretion rate (SMAR), expressed as,
\begin{equation}
\alpha=\frac{dM_*}{dM_\mathrm{h}}\frac{M_\mathrm{h}}{M_*}=\frac{SFR/M_*}{\dot{M_\mathrm{h}}/M_\mathrm{h}}=\frac{\mathrm{sSFR}}{\mathrm{SMAR}}.
\end{equation}
In the \citet{2015ApJ...799...32B}, they assume that $\alpha$ remains constant at $z\geq5$ over the galaxy's star formation history. 
The constant $\alpha$ is motivated because the galaxy's history would be dominated by a single feedback mode; they consider only the stellar feedback (supernovae and reionization) neglecting the AGN feedback. This assumption also implies a relation $M_*\propto M^\alpha_\mathrm{h}$, because equation (4) is interpreted as
\begin{equation}
\alpha=\frac{\frac{d\mathrm{log}M_*}{dt}}{\frac{d\mathrm{log}M_\mathrm{h}}{dt}}=\frac{d\mathrm{log}M_*}{d\mathrm{log}M_\mathrm{h}}.
\end{equation}
Based on this assumption and simulation results for halo
mass function and dark matter accretion rate, \citet{2015ApJ...799...32B} start their calculation from $z=5$ using the abundance matching results for $\alpha$ and the $M_*$-$M_\mathrm{h}$ relation at $z=5$.
Figure \ref{fig_SFRD} shows their SFRDs for $z=5-15$.
Their SFRDs agree well with the observations at $z=5-9$, while they are slightly higher than the recent estimates at $z=10$ based on UV luminosity function observations (Bouwens et al. 2015; Oesch et al. 2018).

On the other hand,  the clustering analysis of $z\sim4-7$ Lyman-break galaxies in \citet{2018PASJ...70S..11H} and \citet{2016ApJ...821..123H} provides an empirical equation for the SFR per halo as
\begin{equation}
SFR=\frac{2\times1.7\times10^{-2}}{(M_\mathrm{h}/M_{\rm br})^{-1.1}+(M_\mathrm{h}/M_{\rm br})^{0.3}}\dot{M_\mathrm{h}},
\end{equation}
where $M_{\rm br}=10^{11.35} M_\odot$.
We use the halo mass function, $\frac{dn}{dM_\mathrm{h}}$, and the dark matter accretion rate, $\dot{M_\mathrm{h}}$, at $z=5-30$ from the Phi-1 simulation in Ishiyama et al. (2015) \footnote{http://hpc.imit.chiba-u.jp/\~\ ishiymtm/db.html}, which is calculated in the redshift range of $z=0-30$ in a $32\ h^{-1}\mathrm{Mpc}$ box with the mass resolution of $3.28\times10^{5}\ M_\odot$.
Hereafter, we call this model HPhi-1.

Figure \ref{fig_SFRD} shows the cosmic SFRDs at $z=5-30$ in the HPhi-1model.
In the redshift range of $z=7-25$, the model is well fitted by a simple power law function,
\begin{equation}
\mathrm{log}(\rho_\mathrm{SFR}/[M_\odot\mathrm{yr^{-1}}\mathrm{Mpc^{-3}}])=-0.32z+0.47.\label{eq_SFRDH18}
\end{equation}
As shown in Figure 1, the HPhi-1 model predicts lower SFRDs than previous models, but the results agree with UV luminosity function observations at $z=5-10$ \citep{2015ApJ...803...34B,2018ApJ...855..105O}.
{For comparison, we also plot the extrapolated fitting formula of \cite{2014ARA&A..52..415M} in Figure 1.
This fitting formula was derived from the observation data at $0<z<8$.
Since the extrapolated fitting formula for $z>8$ is not consistent with the observation results, hereafter, we focus on HPhi-1 model, and \cite{2015ApJ...799...32B} which are based on the halo mass function and  dark matter accretion rate obtained from cosmological simulations.}
\subsection{GRB rate from Pop I\&II stars}

The redshift evolution of the GRB rate does not follow the
star formation rate \citep[e.g.][]{WP10,Lien2014,Lien2015}.
This discrepancy may be explained by the metallicity effect on the progenitor formation
\citep{Yoon2005,Hirschi2005,Woosley2006,Yoon2006,Kinugawa2017b}.
However, we can expect significantly low metallicity at high redshifts,
where its effect may be not important any longer.
Then, the GRB rate will roughly follow the star formation rate differently from the GRB rate at lower redshift..
\citet{WP10} estimated the comoving GRB rate at $z=8$ as $\dot{n}_{\rm GRB}=10^{+12}_{-8.5}~\mbox{Gpc}^{-3}~\mbox{yr}^{-1}$
above the luminosity of $L=10^{50}~\mbox{erg}~\mbox{s}^{-1}$.
Assuming a broken power-law for the redshift evolution,
more recent analysis by \citet{Lien2014,Lien2015} provided
the rate at $z=8$ as $\dot{n}_{\rm GRB}=6.2^{+3.8}_{-5.3}~\mbox{Gpc}^{-3}~\mbox{yr}^{-1}$.

Assuming that the GRB rate is proportional to the star formation rate,
we extrapolate those rates from $z=8$.
Given the field of view $\Omega_{\rm obs}$,
the GRB occurnece probability at $z>z_0$ is
\begin{align}
\frac{dN_{\rm GRB}}{d \Omega_{\rm obs} d t_{\rm obs}}(z>z_0)
=\int_{z_0}^\infty \frac{\dot{n}_{\rm GRB}}{1+z}
\frac{dV_{\rm c}}{dz d\Omega_{\rm obs}} dz,
\end{align}
where $(1+z)$ in the denominator is the time dilation effect
due to the cosmological expansion.
Using the luminosity distance $D_{\rm L}$,
the differential of the comoving volume $V_{\rm c}$ is written as
\begin{align}
\frac{dV_{\rm c}}{dz d\Omega_{\rm obs}}=
\frac{D_{\rm L}^2}{(1+z)^2} \frac{c}{H_0}
\frac{1}{\sqrt{\Omega_{\rm m} (1+z)^3+\Omega_\Lambda}},
\end{align}
where $H_0$, $\Omega_{\rm m}$, and $\Omega_\Lambda$
are the standard cosmological parameters.

The detection rate depends on the GRB spectrum, luminosity function,
and energy band and sensitivity of instruments.
The luminosity functions assumed in \citet{WP10} and \citet{Lien2014,Lien2015}
are broken power-law,
\begin{align}
\frac{d n_{\rm GRB}}{dL} \propto
\left\{
\begin{array}{ll}
\left( \frac{L}{L_\star} \right)^{-a},
& \mbox{for}~L \leq L_\star \\
\left( \frac{L}{L_\star} \right)^{-b},
& \mbox{for}~L > L_\star.
\end{array}
\right.
\label{eq9}
\end{align}

The parameters in \citet{WP10} are $L_\star=10^{52.5}~\mbox{erg}~\mbox{s}^{-1}$,
$a=1.2$, and $b=2.4$, which implies that lower luminosity GRBs
dominate the GRB number. The lower limit in luminosity is taken
as $10^{50}~\mbox{erg}~\mbox{s}^{-1}$.
\citet{Lien2014,Lien2015} adopts $L_\star=10^{52.05}~\mbox{erg}~\mbox{s}^{-1}$,
$a=0.65$, and $b=3.0$.
In this case, GRBs at $L \sim L_\star$ dominate the GRB number.

The GRB spectra are well described by the Band function \citep{Band1993},
whose parameters are the peak energy $\varepsilon_{\rm pk}$,
low-energy index $\alpha$, and high-energy index $\beta$.
While the peak energy in the rest frame is assumed as a constant
$511$ keV in the analysis of \citet{WP10},
\citet{Lien2014,Lien2015} adopts a modified Yonetoku relation
\begin{align}
\varepsilon_{\rm pk}=1.8\times \left(\frac{1}{2.34 \times 10^{-5}}
\times \frac{L}{10^{52}~\mbox{erg}~\mbox{s}^{-1}} \right)^{0.5}~\mbox{keV}.
\end{align}
In this paper, as a typical value, we fix the indices as
$\alpha=-1$ and $\beta=-2.25$.

\begin{figure}[!ht]
	\includegraphics[width=0.5\textwidth,clip=true]{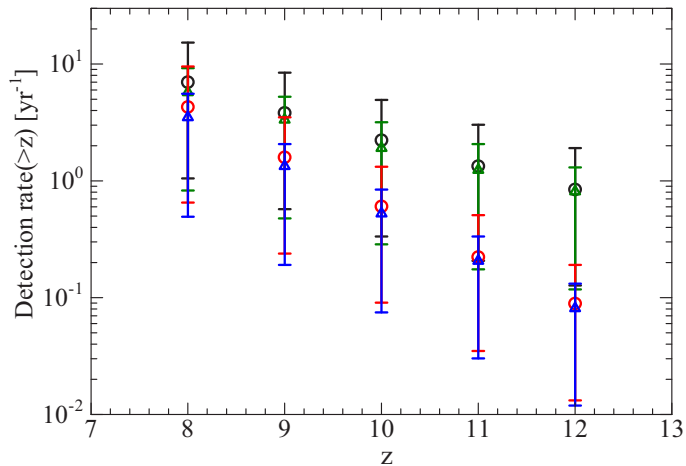}
	\caption{The expectation of the GRB detection with an instrument
		with a sensitivity of $10^{-10}~\mbox{erg}~\mbox{cm}^{-2}~\mbox{s}^{-1}$
		in $0.5$--$4$ keV, and field of view $0.2$ str.
		Black and red circles are estimated with the parameters in \citet{WP10}
		adopting the star formation rate in \citet{2015ApJ...799...32B}
		and HPhi-1, respectively.
		Green and blue triangles are estimated with the parameters in \citet{Lien2015}
		adopting the star formation rate in \citet{2015ApJ...799...32B}
		and HPhi-1, respectively.}
	\label{fig:detectGRB1}
\end{figure}

As a future observation mission, we consider wide field X-ray monitor
with Lobster Eye optics, which may be adopted by the missions in HiZ-GUNDAM
\citep{Yoshida2016}
or THESEUS \citep{YUAN10,Amati2018,Rossi2018}.
With such an instrument, we can expect a sensitivity of
$\sim 10^{-10}~\mbox{erg}~\mbox{cm}^{-2}~\mbox{s}^{-1}$ for 100 s
exposure, and a field of view $\sim 0.2$ str  \footnote{{Private communication with HiZ-GUNDAM working group.
	See also Yuan et al. (2016).} Multiple Lobster Eye systems can enlarge the field of view depending on the budget in future plans. We conservatively assume single Lobster Eye system. }.
In Figure \ref{fig:detectGRB1}, we plot the expectation
of the detection rate adopting the two models:
the models in \citet{WP10} (circles) and \citet{Lien2014,Lien2015} (triangles) for
the GRB rate, luminosity function and spectral peak energy.
If we adopt the star formation rate in \citet{2015ApJ...799...32B} (black and green),
both the models suggest a few GRB detection per year for $z>10$.
However, the Hphi-1 SFR,
which seems consistent with the observed rate at $z \simeq 10$,
leads to a detection rate $\ll 1$ for $z>12$ (see red and blue symbols).
In addition, we should take into account the efficiency
of the redshift confirmation, which depends on the performance
of the follow-up infrared telescope onboard HiZ-GUNDAM or THESEUS.
The confirmation of GRBs at $z>12$ seems not easy.
However, in other words, this provides the opportunity to detect
other types of trangient phenomena at high redshifts,
such as GRBs from Pop III stars.

\section{GRB from Pop III stars}\label{GRBfromPopIII}

\subsection{Pop III star formation rate}
{Pop III stars are first stars formed from  metal-free gases. Pop III stars are
 massive \cite[e.g.][]{McKee2008,Hosokawa2011}, and have no stellar wind mass loss\citep{Krticka2006}.
Since Pop III stars are formed in anomalous circumstance, their formation history may be different from Pop I\&II stars.}

At present, we have not significant constraint on the Pop III star formation rate from observations.
However,  the Pop III star formation rate has been estimated using the cosmological simulation.
We consider two Pop III SFRDs in \cite{de Souza2011} and \cite{Inayoshi2016}.

The SFRD of \cite{de Souza2011} is calculated by a semi-analytical approach, in which they assume that Pop III stars are formed in  dark matter haloes at their collapse. They adopt the Sheth-Tormen mass function \citep{Sheth1999} to estimate the number of dark matter halos at given redshift. They divide the populations into the Pop III.1 and Pop III.2 stars.
Pop III.1 stars are the first generation Pop III stars formed in the dark matter 'minihaloes', where only H$_2$ molecular cooling is the dominant cooling process.
Pop III.2 stars are the second generation Pop III stars formed from ionized gases that are in the H$_{\rm II}$ region made by previous Pop III stars or a virialization shock in the halo with the virial temperature $T_{\rm Vir}\gtrsim 10^4$ K.
The hydrogen deuteride (HD) cooling is efficient below 200 K in this region due to an enhanced free electron fraction. 
As a result, the typical mass of Pop III.2 is slightly smaller than the Pop III.1 \citep{Hosokawa2012} but still massive enough to form a BH at their end of life, which may launch a GRB jet.

In order to determine how many Pop III.1 and Pop III.2 stars are formed in collapsed haloes, \cite{de Souza2011} introduce a parameter, the star formation efficiency $f_{*}$.
The original SFRD in \cite{de Souza2011} are obtained with $f_*=0.1$, and 0.01 for Pop III.1, and Pop III.2, respectively.
In de Souza et al (2011), using a criteria based on the virial mass and reionization history, the evolutions of SFRD of Pop III.1 and PopIII.2 are calculated. They also follows the progress of the metal enrichment by protogalactic wind, which prevents the Pop III star formation.

The SFRD in \cite{Inayoshi2016} is also calculated by a semi-analytical approach using the same Sheth-Tormen mass function \citep{Sheth1999} based on the model described in \cite{Visbal2015}.
The treatments for the progresses of ionization and metal enrichment are different from the model in \cite{de Souza2011}.
\cite{Inayoshi2016} take into account the constraint of the star formation from the Thomson scattering optical depth $\tau_{\rm e}=0.066+1\sigma$ where $\sigma=0.016$, which is measured by \cite{Planck2016a}. These changes generate significant constraint on models of Pop III star formation \citep{Visbal2015, Inayoshi2016}, although this constraint depends on some parameters such as the escape fraction of photon, the initial mass function (IMF), and so on.  \cite{Inayoshi2016} shows that the constraint of the total Pop III star formation density is $\rho_{*,III}\lesssim6\times10^5~\msun\rm~Mpc^{-3}$  for $\tau_{\rm e}=0.066+1\sigma$, the escape fraction of ionizing photons from mini-halos $f_{\rm esc}=0.1$, and the flat IMF ($10~\msun<M<100~\msun$).

\begin{figure}[!ht]
	\includegraphics[width=0.5\textwidth,clip=true,clip]{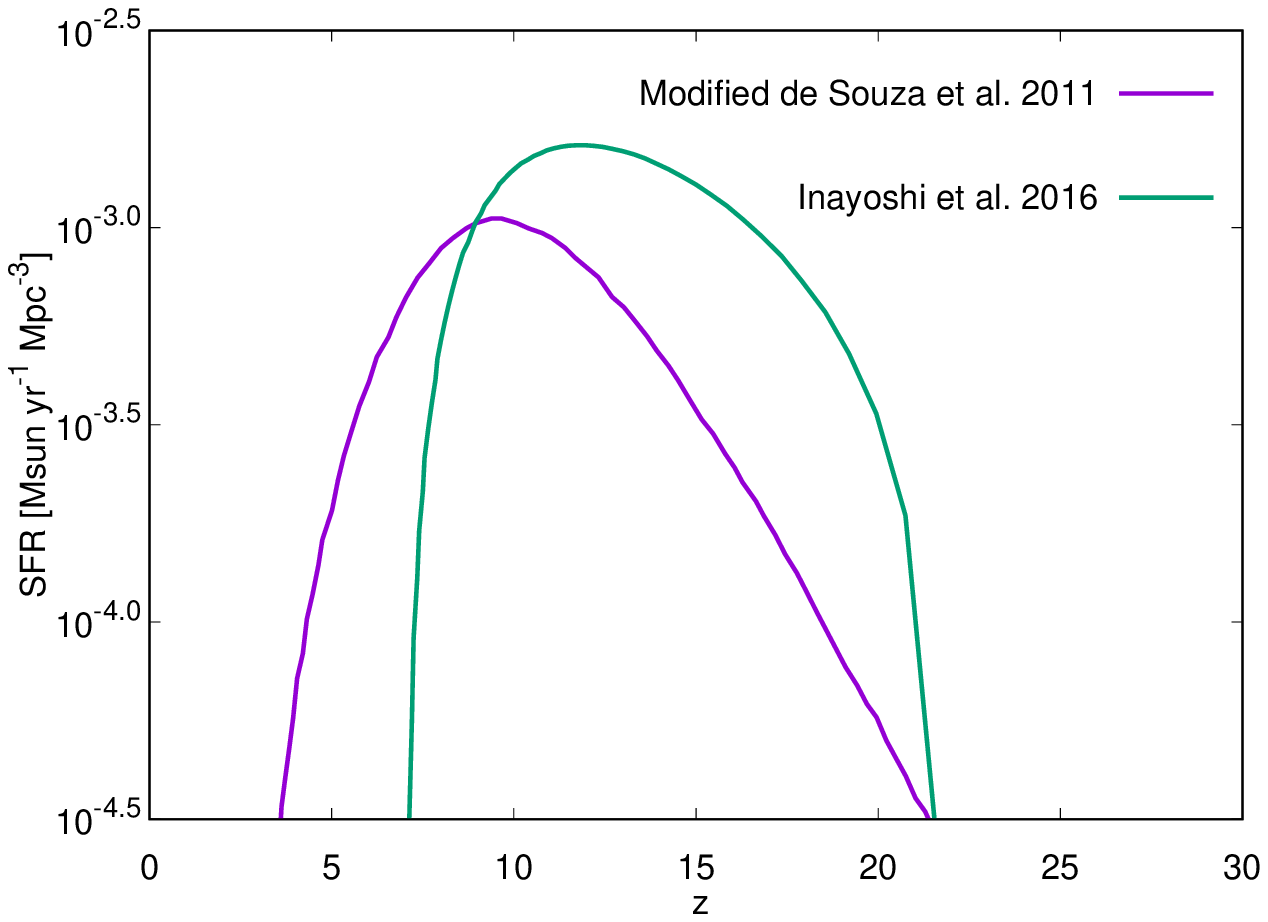}
	\caption{The SFRD of \cite{de Souza2011} modified with the constraint of $\rho_{*,III}\lesssim6\times10^5~\msun\rm~Mpc^{-3}$ , and the SFRD of \cite{Inayoshi2016}}
	\label{fig:PopIIISFR}
\end{figure}

The SFRD in \citet{de Souza2011} already conflicts with the limit
by the Planck observation so that we adopt the SFRD in \citet{de Souza2011} decreasing
 by a factor of 0.3, which corresponds to the upper-limit
of $\rho_{*,III}$ given  by \citet{Inayoshi2016}.
Figure \ref{fig:PopIIISFR} shows the SFRDs we adopt in this paper based on \cite{de Souza2011}  and \cite{Inayoshi2016}.

\subsection{Ultra-long GRB rate from Pop III}

{Hereafter, we assume that the GRB rate is proportional to SFRDs in Figure 3. The number of stars are calculated from
the initial mass function, which is assumed as
 the flat
\begin{align}
\frac{dN_{\mbox{III}}}{dM}=\mbox{const.}
\end{align}
between $10 M_\odot$ and $100 M_\odot$.
This implies that the average mass of Pop III stars, $M_{\mbox{III}}$, is $55 M_\odot$.}

In the most optimistic scenario, all such heavy stars give rise to a GRB.
Then, the apparent GRB rate is simply estimated as $f_{\rm B} SFR/M_{\mbox{III}}$,
where $f_{\rm B}$ is the beaming factor.
For metal free stars like Pop III stars, however, the stellar wind is suppressed so that
a massive envelope remains at the collapse \citep{Krticka2006}.
In such cases, its long free fall time leads to ultra-long GRBs \citep{Nakauchi2012,Nakauchi2013},
whose duration is $\sim 10^4$ s.
The jet opening angles in ultra-long GRBs are estimated as wider than
$10^\circ$ \citep{Levan2014}.
Here, we adopt an optimistic opening angle $\theta_{\rm j}=20^\circ$,
which implies the beaming factor $f_{\rm B}=0.06$.
{Finally we obtain the comoving GRB rates at $z=8$
as $\dot{n}_{\rm GRB}=1000~\mbox{Gpc}^{-3}~\mbox{yr}^{-1}$
and $440~\mbox{Gpc}^{-3}~\mbox{yr}^{-1}$ for the SFRs in \citet{de Souza2011}
and \citet{Inayoshi2016}, respectively.}

\begin{figure}[!ht]
	\includegraphics[width=0.5\textwidth,clip=true]{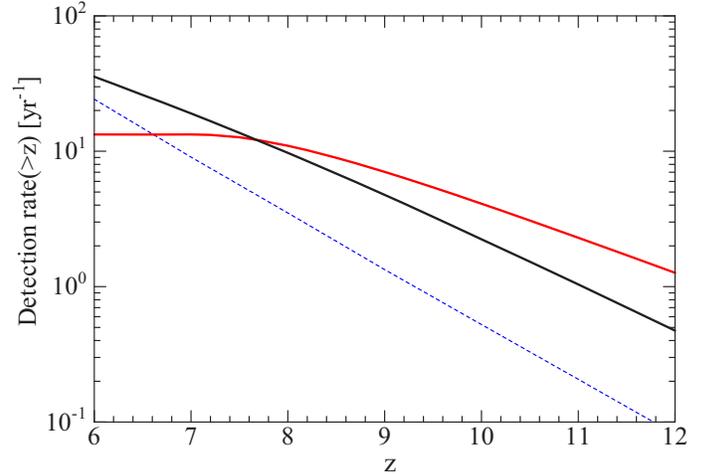}
	\caption{The expectation of the ultra-long GRB detection
		originated from Pop III stars with an instrument
		with a sensitivity of $2 \times 10^{-11}~\mbox{erg}~\mbox{cm}^{-2}~\mbox{s}^{-1}$
		in $0.5$--$4$ keV, and field of view $0.2$ str.
		The black and red lines are detection rates with SFRs of modified one in \citet{de Souza2011} and \citet{Inayoshi2016},
		respectively. The blue dashed line is the fiducial detection rate
		for Pop I\&II GRBs with the HPhi-1 star formation rate
		taken from Figure \ref{fig:detectGRB1}.}
	\label{fig:detectGRB2}
\end{figure}

The observed typical luminosity of ultra-long GRBs is
$\sim 10^{49}~\mbox{erg}~\mbox{s}^{-1}$ \citep{Gendre2013,Peng2013,Levan2014}.
As the luminosity function in equation (\ref{eq9}),
we adopt the same parameters with those in \citet{Lien2014,Lien2015}
but $L_\star=10^{49}~\mbox{erg}~\mbox{s}^{-1}$ with
the lower and upper limits $10^{47}~\mbox{erg}~\mbox{s}^{-1}$ and
$10^{51}~\mbox{erg}~\mbox{s}^{-1}$, respectively.
The GRB spectra are assumed as the same Band function with the modified Yonetoku
relation.
The integration time for an instrument like HiZ-GUNDAM or THESEUS on geocentric orbits
would be limitted below $\sim 1000$ s.
Assuming a sensitivity of
$2 \times 10^{-11}~\mbox{erg}~\mbox{cm}^{-2}~\mbox{s}^{-1}$ for 1000 s
exposure between 0.5 and 4 keV, and a field of view $\sim 0.2$ str\footnote{{Private communication with HiZ-GUNDAM working group.
	See also Yuan et al. (2016). }},
we plot the expectation of the ultra-long GRB detection in Figure \ref{fig:detectGRB2}.

The estimated detection rates of ultra-long GRBs
are significantly higher than the usual GRB rate.
The dominant sources in the HiZ-GUNDAM/THESEUS era may be
ultra-long GRBs originated from Pop III stars.
However, the assumption that all Pop III stars die as a GRB
may be too optimistic.
Practically the rates in Figure 4 are upper limits of Pop III ultra-long GRB rates. 
{In this optimistic case, ultra-long GRBs from Pop III stars may be detected even at $z=6-8$ especially for the modified de Souza SFRD model.}
If more than 10\% of Pop III stars induce an ultra-long GRB, $\sim$ one detection per year is expected.
\citet{de Souza2011} assumed that only 1\% of Pop III stars
cause a GRB at their end of life.
Under such a conservative assumption, detection of an ultra-long GRB seems
very difficult.

\subsection{Pop III binary population synthesis \& binary merger model}\label{PopIIIbinary}
\begin{table*}
	\caption{The initial distribution functions in This paper.
	}
	\label{IDF}
	\begin{center}
		\begin{tabular}{cccc}
			\hline
			IMF	& Initial Mass Ratio Function & 	Initial Period Function  &Initial Eccentricity function\\
			\hline
			flat  & flat & 1/a & e\\
			$10~\msun<M_1<100~\msun$&$10~\msun/M_1<M_2/M_1<1$&$a_{\rm min}$*$<a<10^6~\rsun$&$0<e<1$\\
			\hline
		\end{tabular}\\
		{* We choose $a_{min}$ as the minimum separation when the binary does not fulfill the Roche lobe \citep{Kinugawa2014}.}
	\end{center}
\end{table*}

Classical long GRBs, whose typical luminosity is more than $10^{52}~\mbox{erg}~\mbox{s}^{-1}$ with the typical duration of $10~\mbox{s}$, are favorable to detect rather than ultra-long GRBs. The binary interaction may produce ideal progenitors to cause classical GRBs.

In order for long GRBs to occur, the progenitors need high angular momentum.
Although the angular momentum of Pop III stars is unknown, the remnant of the binary merger during a common envelope (CE) phase possibly have a high angular momentum.
{When the radius of primary giant suddenly becomes larger or a radical mass transfer shrinks the orbit, the secondary star sometime plunges into the primary envelope. The secondary star spirals in and the envelope of primary will be evaporated. After the CE phase, the binary becomes a close binary which consists of the secondary and the core of the primary giant or the two stars merges}
{ during a CE phase. In the latter case, the envelope evaporated, and a highly rotating 
helium star would remain \citep{Fryer2005}.
Furthermore, the highly spinning progenitors evolve as chemically homogeneous stars \citep{Yoon2012}.
Since such highly rotating stars have small radius, the jet can break out the stellar surface with a high accretion rate like Pop I\&II GRBs.
For such idealized progenitors, GRBs can be as luminous as usual observed GRBs
with duration of $\sim$ 10 s \citep[e.g.][]{Suwa2011}.}
Thus, we focus on the binary merger model \citep{Fryer2005} as the Pop III GRB progenitor.
We consider two channels for Pop III GRB progenitors: (1) highly rotating helium stars and (2) highly rotating main-sequence stars.

The highly rotating helium stars are made by the binary mergers during a CE phase that contains only post main sequence stars.
On the other hand, highly rotating main-sequence stars are made by the binary mergers during a CE phase that contains a Giant star and a main sequence star.
Using the population synthesis method, we calculate these binary merger fraction of Pop III stars and estimate the Pop III long GRB rate.
{According to the binary population synthesis method, we set the zero age main sequence binary parameters, such as primary mass $M_1$, mass ratio $M_2/M_1$, separation $a$, and eccentricity $e$, using the initial distribution functions, and calculate each stellar evolution. The numerical code judges whether stars experience the binary interactions (BIs) such as the tidal friction, the mass transfer, the CE phase, and so on and updates the parameters $M_1, M_2, a$, and $e$ in each time step. We repeat this calculation using different initial binary parameters chosen by the Monte-Carlo method \citep{Kinugawa2014}.}
{	We use the flat IMF that is suggested by some simulations {\citep{Hirano2014,Susa2014}.}
	We assume the other initial distributions are the same as those of Pop I binaries {\citep{Heggie1975,Abt1983,Kobulnicky2007} as summarized in Table 1..}
	Using the Monte Carlo method with those initial distribution functions, we calculate the Pop III evolutions of the stellar radius and the core mass and check whether the binary interaction occurs or not.}
We calculate $10^6$ binaries for each models.

The calculation code is the same as the Pop III binary population synthesis code in \cite{Kinugawa2014} and \cite{Kinugawa2017}.
{This code was used to calculate the binary black hole merger rate and the detection rate of LIGO gravitational wave observations.
The binary black hole merger rate calculated by this code \citep{Kinugawa2014,Kinugawa2016} with a similar initial parameter set to
that in this paper is consistent with the LIGO result \citep{Abbot2016b,Abbot2018}.
}

\begin{figure}[!ht]
	\includegraphics[width=0.5\textwidth,clip=true]{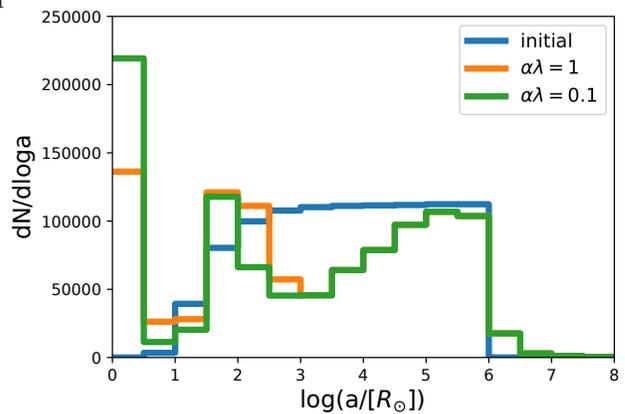}
	\caption{The blue line is the initial separation distribution for $10^6$ binaries. Orange and green lines are final separation distributions after binary stars' death  for $\alpha_{\rm CE}\lambda=1$ and $\alpha_{\rm CE}\lambda=0.1$ cases, respectively.}
	\label{CEsep}
\end{figure}

{We use the following formalism for the CE phase.
	The criterion of the mass transfer  leading to a CE phase is the same as that of our previous paper \citep{Kinugawa2014}.
	In order to calculate the separation just after the CE phase $a_{\rm f}$, we use the energy balance prescription  \citep{Webbink1984}}
\begin{equation}
\alpha_{\rm CE}\left(\frac{GM_{\rm{c,1}}M_2}{2a_{\rm{f}}}-\frac{GM_1M_2}{2a_{\rm{i}}}\right)=\frac{GM_{\rm{1}}M_{\rm{env,1}}}{\lambda R_1},
\label{eq:ce1}
\end{equation} 
for a binary of a giant star and
a main sequence star, where $a_{\rm i}$, $R_1$, $M_1$, $M_{\rm c,1}$, $M_{\rm env,1}=M_1-M_{\rm c,1}$, and $M_2$ are the binary separation just before the CE phase, the radius, the mass, the core mass and the envelope mass of the giant, and the mass of the companion star, respectively.
The value {$\alpha_{\rm CE}$ is the efficiency parameter how much the orbital energy can be used in ejecting the envelope. The parameter $\lambda$ is for the envelope binding energy.}
{If the companion star is also a giant, Equation (\ref{eq:ce1}) changes into
	\begin{align}
		\alpha_{\rm CE}\left(\frac{GM_{\rm{c,1}}M_{c,2}}{2a_{\rm{f}}}-\frac{GM_1M_2}{2a_{\rm{i}}}\right)=&\frac{GM_{\rm{1}}M_{\rm{env,1}}}{\lambda R_1}\notag\\
	&+\frac{GM_{\rm{2}}M_{\rm{env,2}}}{\lambda R_2},
	\label{eq:ce2}
	\end{align}
	where  $M_{\rm c,2}$, $M_{\rm env,2}=M_2-M_{\rm c,2}$, and  $R_2$ are  the core mass, the envelope mass, and the radius of the companion star, respectively \citep{Dewi2006}.} 
The CE parameters $\alpha_{\rm CE}$ and $\lambda$ are not well understood \citep{Ivanova2013}.
{We adopt the typical CE parameter values adopted in the previous binary population studies ($\alpha_{\rm CE}\lambda=1$ and $0.1$) \citep{Belczynski2007,Kinugawa2014}.
The simulation of the CE phase is so difficult that the CE parameter are theoretically uncertain \citep[e.g.][]{Ivanova2013}, but there are some observation constraints.
Those values ($\alpha_{\rm CE}\lambda=1$ and $0.1$) can reproduce the observation results such as the separation distribution of observed white dwarf binaries,
and the period-eccentricity distribution of observed binary pulsars \citep[e.g.][]{Zorotovic2010,Hijikawa2019}.
The Pop III binary black hole merger rates using those values are consistent to the LIGO's result \citep{Kinugawa2016,Abbot2016b}.}
	A smaller $\alpha_{\rm CE} \lambda$ leads to a closer separation after the CE phase, and vice versa.
	Thus, the smaller $\alpha_{\rm CE}\lambda$ implies an efficient stellar merger.{Figure \ref{CEsep} shows the $\alpha_{\rm CE}\lambda$ dependence of the binary separation.
This figure demonstrates that a smaller $\alpha_{\rm CE}\lambda$ makes binaries easier to merge. }
	
{Just after the CE phase, if $a_{\rm f}$ is less than the sum of the radius of the giant's core and the radius of the companion star {(or the sum of radii of the two giants' cores if the companion star is also a giant)}, we assume that the binary stars merge.}

If two giants merge during the CE phase,  the merged helium star obtains a large angular momentum from the orbital angular momentum of the binary.
Thus, we assume that the helium star has a spin angular momentum of the Kepler velocity.
The mass of the highly rotating helium star is the sum of the primary helium core and the secondary helium core.
The radius  $R_{\rm rem}$ is calculated by Equation (81) in \cite{Hurley2000}.
{The spin angular momentum is calculated by $kMR^2\Omega_{k}$, where $k$ is a parameter for the momentum of inertia which is calculated from the density profile. We assume that $k$ of the highly rotating helium star is the same as the value adopted in \cite{Hurley2000} for the helium dense convective core of giants ($k=0.21$).}
{After the merger, the highly rotating helium star loses the angular momentum by the stellar wind mass loss.}
The stellar wind mass lose rate of Pop III stars is very weak ($\sim10^{-14}\msun~\rm{yr}^{-1}$).
But, in the case of highly rotating stars, the rotation effect enhances the mass loss rate.
We use the following formula as the mass loss rate for the rotating helium stars,
\begin{equation}
\dot{M}={\rm min}\left(\frac{3}{10}\frac{M}{\tau_{\rm KH}},~10^{-14}\left(1-\frac{\Omega}{\Omega_{\rm K}}\right)^{-0.43}\right)
\end{equation}  
{\citep{Yoon2012}, where $\tau_{\rm KH}=GM^2/RL$, $M$, $R$, $L$, $\Omega$, and $\Omega_{\rm K}$ are the Kelvin-Helmholtz timescale, the stellar mass, the stellar radius, the stellar luminosity, the angular velocity of the star, and the angular velocity of the Kepler rotation, respectively.}
The angular momentum loss due to the wind mass loss is written as
\begin{equation}
\dot{J}=\frac{2}{3}\dot{M}R^2\Omega.
\end{equation}
{We assume that the highly rotating helium stars evolve as the chemically homogeneous stars \citep{Maeder1987} and will change into a CO star.}
{In the case of the chemically homogeneous stellar evolution, the heavy elements are possibly carried to the stellar surface. But, we do not consider the effect of the surface heavy elements on the mass  loss enhancement, because \cite{Krticka2009} and \cite{Muijres2012} show that such effect is rather moderate.}
When stars collapse, we treat them as a direct collapse.
If the mass of the star after the collapse is larger than 3 $ M_{\odot}$, the star is regarded as a BH. 
If the mass of helium stars is more massive than 60 $\msun$,  the stars possibly cause pair instability supernovae \citep{Woosley2007}.
Thus, we assume highly rotating helium stars whose masses are $3~\msun<M<60~\msun$ as GRB progenitors.

{On the other hand, if a giant and a main sequence star merge during the CE phase, the merged remnant becomes a highly rotating main-sequence star.
	{The mass of the highly rotating main-sequence star is the sum of the primary giant's core and the secondary main-sequence star.}
We use the results of \cite{Yoon2012} to determine the fate of highly rotating main-sequence stars.
	\cite{Yoon2012} shows that the highly rotating main-sequence star whose mass is $13~\msun\lesssim M$ can evolve as the chemically homogeneous, and if their mass is $\lesssim 84~\msun$, they do not become a pair instability supernova, and the inner cores of those stars have a significantly high angular momentum to launch a GRB jet.
	Thus, we assume highly rotating main-sequence stars whose mass range is $13~\msun< M<84~\msun$ become long GRB progenitors \citep{Yoon2012}.
}


\subsection{Classical long GRB rate from Pop III}
Table \ref{PopIIIGRB} shows the numbers of the long GRB progenitors for $10^6$ binaries, obtained from the calculation shown in the section \ref{PopIIIbinary}.
A few percents of Pop III binaries can cause classical GRBs.
We calculate the long GRB rate $R_{\rm GRB}$ from Pop III stars, using the beaming factor $f_{\rm B}=0.01$, the binary fraction $f_{\rm b}=0.5$, Pop III SFRs, and the long GRB fraction of Pop III $f_{\rm GRB}$ which consist of highly rotating Helium stars and highly rotating main-sequence stars as
\begin{equation}
R_{\rm GRB}=f_{\rm B}\cdot f_{\rm GRB}\cdot\left(\frac{f_{\rm b}}{1+f_{\rm b}}\right)\cdot \frac{SFR}{M_{\rm III}} .
\end{equation}
{The beaming factor $f_{\rm B} = 0.01$ is chosen to make the opening	angle the same order of the typical value for the Pop I\&II case \citep{Liang2008}. The binary fraction $f_{\rm b}=0.5$ is also the same as those in the Pop I\&II case \citep{Sana2012, Sana2013}, which is consistent with the BH-BHS merger rate \cite{Kinugawa2016,Belczynski2016}.}

\begin{table*}[!ht]
	\caption{The long GRB fraction of Pop III $f_{\rm GRB}$ which consist of highly rotating Helium stars and highly rotating main-sequence stars}
	\label{PopIIIGRB}
	\begin{center}
		\begin{tabular}{ccc}
			\hline
			progenitor type	& Highly rotating Helium stars & highly rotating main-sequence stars  \\
			\hline
			$\alpha_{\rm CE}\lambda=1$ model& 1.1\% & 2.5\% \\
			$\alpha_{\rm CE}\lambda=0.1$ model & 1.6\%&7.8\%  \\
			
			\hline
		\end{tabular}\\
	\end{center}
\end{table*}
\begin{figure}[!ht]
	\includegraphics[width=0.5\textwidth,clip=true]{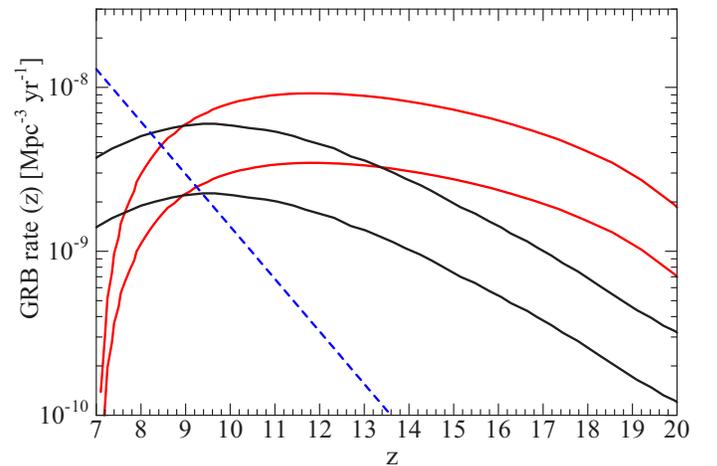}
	\caption{The classical long GRB rate from Pop III stars. The black and red lines are detection rates
		with SFRs modified one in de Souza et al.
		(2011) and Inayoshi et al. (2016), respectively.
		The upper and lower lines correspond to the parameter of $\alpha \lambda=0.1$
		and $1$, respectively. The blue dashed
		line is the fiducial detection rate for Pop I\&II GRBs with the Hphi-1 model taken from Figure 2.}
	\label{fig:5}
\end{figure}
\begin{figure}[!ht]
	\includegraphics[width=0.5\textwidth,clip=true]{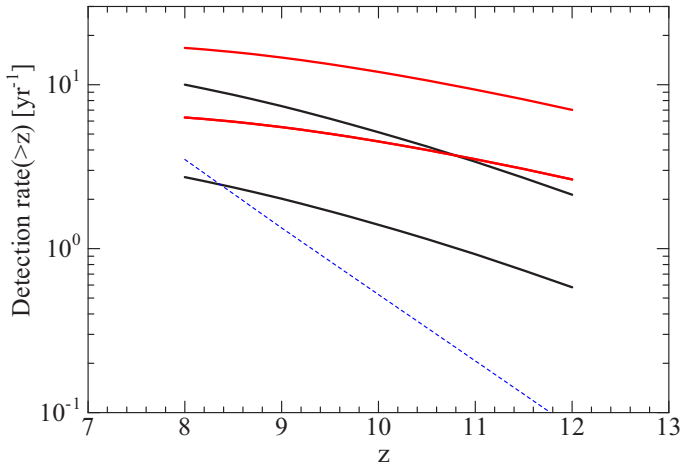}
	\caption{The expectation of the classical GRB detection originated
		from Pop III stars based on the binary interaction model
		with an instrument with a sensitivity of
		$10^{-10}~\mbox{erg}~\mbox{cm}^2~\mbox{s}^{-1}$ in 0.5--4 keV, and a field of view 0.2 str. The black and red lines are detection rates
		with SFRs modified one in de Souza et al.
		(2011) and Inayoshi et al. (2016), respectively.
		The upper and lower lines correspond to the parameter of $\alpha \lambda=0.1$
		and $1$, respectively.
		The blue dashed
		line is the fiducial detection rate for Pop I\&II GRBs with the Hphi-1 model taken from Figure 2.}
	\label{fig:PopIIIlongGRB}
\end{figure}
{Figure \ref{fig:5} shows the classical long GRB rate from Pop III stars.}
Figure \ref{fig:PopIIIlongGRB} shows the expectation of the classical GRB detection originated
from Pop III stars based on the binary interaction model
with an instrument with a sensitivity of
$10^{-10}~\mbox{erg}~\mbox{cm}^2~\mbox{s}^{-1}$ in 0.5--4 keV, and a field of view 0.2 str.
{The same parameters as those in Lien et al. (2014) are adopted as the classical GRB luminosity function, and the modified Yonetoku relation is used for the spectral peak energy.}
The black and red lines are detection rates
with SFR in de Souza et al.
(2011) and Inayoshi et al. (2016), respectively.
The upper and lower lines correspond to the parameter of $\alpha_{\rm CE} \lambda=0.1$
and $1$, respectively.
The blue dashed
line is the fiducial detection rate for Pop I\&II GRBs based on the HPhi-1 model taken from Figure 2.

Although the GRB fraction is small compared to the assumption for ultra-long GRBs in section 3.2, the brighter luminosity function provides higher detection rates as shown in Figure 5. 
The SFRDs for Pop III stars assumed in this paper, which do not violate the constraints given by the Planck observation, imply higher detection rates for Pop III GRBs than the rate for Pop I\&II GRBs. This is encouraging for the future observational missions such as HiZ-GUNDAM or THESEUS.

\section{Conclusion and Discussion}
The Hphi-1 model suggests that the SFRD calculated from clustering analysis of galaxies and UV luminosity function observations steeply decreases at high redshift compared to the extrapolated SFRD of the {\cite{2014ARA&A..52..415M} model}.
This result shows that the Pop I\&II stars are hard to contribute for long GRBs at high redshift. However, the SFRD of Pop III stars can be higher than Pop I\&II SFRD.
{At $z=8-9$, the detectable Pop I\&II GRB rate based on HPhi-1+Lien et al. (2014) model is 136 $\rm yr^{-1}$ in the whole sky.
	On the other hand, the Pop III GRB rate using Inayoshi et al. (2016) SFRD is 50 $\rm yr^{-1}$.
	At $z=9-10$, the Pop III GRB rate (63 $\rm yr^{-1}$) is  almost the same as the  Pop I\&II GRB rate (51 $\rm yr^{-1}$).
	GRB events at $z=8-9$ like GRB090423 \citep{Chandra2010} and GRB090429B \citep{Cucchiara2011} would be Pop III GRBs with a probability of a few tens of percent.
	However, we have not found Pop III-like signature for GRB090423 and GRB090429B at present. Note that our binary merger model for Pop III GRBs yields classical GRBs, whose characteristic may be similar to other usual long GRBs.}

In this paper, we consider the GRB from Pop III stars, using two SFRDs of Pop IIII considering the constraint from the Planck observation.
We calculate the detection rate of Pop III GRBs by future observations such as HiZ-GUNDAM and THESEUS.
In the pessimistic model, since the Pop III stars hold the hydrogen envelope because of the weak stellar wind, the Pop III stars are hard to launch a classical long GRB.
In this case, Pop III stars might launch an ultra-long GRB. Only if more than 10\% of Pop III stars launch a GRB jet, the future missions can detect such an ultra-long GRB per year.

However,  many massive binary black holes confirmed by gravitational waves \citep{Abbot2018} might be remnants  of Pop III binaries \citep{Kinugawa2016}.
If a significant fraction of Pop III stars are formed as a binary, we expect that  highly rotating helium stars and highly rotating main-sequence stars are formed via the binary merger, and they evolve as a chemically homogeneous star.   
Our population synthesis calculation shows that several \% of Pop III binaries become such highly rotating stars which possibly launch a long GRB. 
On the other hand, \cite{Belczynski2007} consider Pop III GRB progenitors which lose the envelope and spin up due to tidal spin up and show that such progenitors may be a very small fraction ($\lesssim1\%$). 
Thus, the binary merger is more effective process to make a Pop III GRB progenitor than the tidal spin up.
If such highly rotating stars launch a classical GRB resembling low redshift long GRBs, they can be detected by HiZ-GUNDAM and THESEUS.
Those future observations help us reveal the Pop III SFRD.

\section*{Acknowledgment}
We appreciate D. Yonetoku and T. Ishiyama for the information on the instruments for HiZ-GUNDAM and for the data of N-body simulations, respectively.
This work was supported by JSPS KAKENHI Grant Number 18J00558(TK), 16J03329(YH), 16K05291, and 18K03665 (KA). 
This work is carried out by the joint research program of the Institute for
Cosmic Ray Research (ICRR), The University of Tokyo.


\end{document}